\documentclass{aastex62}

\received{...}
\revised{...}
\accepted{...}

\submitjournal{ApJ}

\shorttitle{Accelerated magnetic reconnection}
\shortauthors{Li et al.}

\begin{document}

\title{Magnetic reconnection between loops accelerated by a nearby filament eruption}

\correspondingauthor{Leping Li}
\email{lepingli@nao.cas.cn}

\author[0000-0001-5776-056X]{Leping Li}
\affil{CAS Key Laboratory of Solar Activity, National Astronomical Observatories, Chinese Academy of Sciences, Beijing 100101, People's Republic of China}
\affiliation{University of Chinese Academy of Sciences, Beijing 100049, People's Republic of China}

\author{Hardi Peter}
\affiliation{Max Planck Institute for Solar System Research, 37077 G\"{o}ttingen, Germany}

\author[0000-0002-9270-6785]{Lakshmi Pradeep Chitta}
\affiliation{Max Planck Institute for Solar System Research, 37077 G\"{o}ttingen, Germany}

\author[0000-0001-5705-661X]{Hongqiang Song}
\affiliation{Shandong Provincial Key Laboratory of Optical Astronomy and Solar-Terrestrial Environment, and Institute of Space Sciences, Shandong University, Weihai, Shandong 264209, People's Republic of China}

\author{Kaifan Ji}
\affiliation{Yunnan Observatories, Chinese Academy of Sciences, 650216 Kunming, People's Republic of China}

\author{YongYuan Xiang}
\affiliation{Yunnan Observatories, Chinese Academy of Sciences, 650216 Kunming, People's Republic of China}

\begin{abstract}

Magnetic reconnection modulated by non-local disturbances in the solar atmosphere has been investigated theoretically, but rarely observed. In this study, employing H$\alpha$ and extreme ultraviolet (EUV) images and line of sight magnetograms, we report acceleration of reconnection by adjacent filament eruption. In H$\alpha$ images, four groups of chromospheric fibrils are observed to form a saddle-like structure. Among them, two groups of fibrils converge and reconnect. Two newly reconnected fibrils then form, and retract away from the reconnection region. In EUV images, similar structures and evolution of coronal loops are identified. Current sheet forms repeatedly at the interface of reconnecting loops, with width and length of 1-2 and 5.3-7.2\,Mm, and reconnection rate of 0.18-0.3. It appears in the EUV low-temperature channels, with average differential emission measure (DEM) weighed temperature and EM of 2\,MK and 2.5$\times$10$^{27}$\,cm$^{-5}$. Plasmoids appear in the current sheet and propagate along it, and then further along the reconnection loops. The filament, located at the southeast of reconnection region, erupts, and pushes away the loops covering the reconnection region. Thereafter, the current sheet has width and length of 2 and 3.5\,Mm, and reconnection rate of 0.57. It becomes much brighter, and appears in the EUV high-temperature channels, with average DEM-weighed temperature and EM of 5.5\,MK and 1.7$\times$10$^{28}$\,cm$^{-5}$. In the current sheet, more hotter plasmoids form. More thermal and kinetic energy is hence converted. These results suggest that the reconnection is significantly accelerated by the propagating disturbance caused by the nearby filament eruption.

\end{abstract}


\section{Introduction} \label{sec:int}

As the reconfiguration of magnetic field geometry, magnetic reconnection plays an elemental role in magnetized plasma systems, e.g., the solar and stellar coronae and planetary magnetospheres, throughout the university \citep{2000mrmt.conf.....P}. It is used to explain the release of magnetic energy and its conversion to other forms, such as thermal and kinetic energy \citep{2010RvMP...82..603Y}. In solar physics, numerous theoretical studies of magnetic reconnection have been undertaken to explain various solar activities, such as flares, filament eruptions, coronal mass ejections, and jets \citep{1999Ap&SS.264..129S, 2000JGR...105.2375L, 2011LRSP....8....1C}. In the two-dimensional (2D) models, magnetic reconnection takes place at an X-point where anti-parallel magnetic field lines converge and reconnect \citep{2000mrmt.conf.....P, 2010RvMP...82..603Y, 2020PRSA...476.20190867N}. However, the process of magnetic reconnection is difficult to observe directly. 

In the solar corona, magnetic flux is frozen into the coronal plasma \citep{2014masu.book.....P}. The coronal structures, e.g., loops and filament threads, and their structural changes thus outline the magnetic field topology and its evolution. Using the remote-sensing observations, many signatures of magnetic reconnection have been reported. These include reconnection inflows \citep{2001ApJ...546L..69Y, 2005ApJ...622.1251L, 2009ApJ...703..877L}, current sheets \citep{2008ApJ...686.1372C, 2010ApJ...723L..28L, 2012SoPh..276..261S, 2016NatPh..12..847L, 2016ApJ...829L..33L, 2018ApJ...866...64C, 2019ApJ...874..146H}, reconnection outflows \citep{2012ApJ...745L...6T, 2014ApJ...797L..14T, 2016ApJ...818L..27C, 2016Ap&SS.361..301L}, plasmoid ejections \citep{2012PhRvX...2b1015S, 2013A&A...557A.115K, 2019A&A...628A...8P, 2020A&A...633A.121X}, loop-top hard X-ray sources \citep{1994Natur.371..495M, 2013NatPh...9..489S}, supra-arcade downflows \citep{2000SoPh..195..381M, 2003SoPh..217..267I, 2016ApJ...829L..33L}, cusp-shaped post-flare loops \citep{1992PASJ...44L..63T, 2018ApJ...853L..18Y}, and coronal structural reconfigurations \citep{2013ApJ...776...57Z, 2014A&A...570A..93L, 2018ApJ...864L...4L, 2018ApJ...868L..33L, 2019ApJ...884...34L, 2018ApJ...863L..22K}.

In the solar chromosphere, observational evidence of magnetic reconnection has also been presented \citep{2020scts1143101914636Y}. Using the H$\alpha$ images from the New Vacuum Solar Telescope \citep[NVST;][]{2014RAA....14..705L} with high spatial and temporal resolution, signatures of magnetic reconnection are observed at an X-type configuration between a pair of interacting fibrils \citep{2015ApJ...798L..11Y, 2016ApJ...819L..24Y}. During the process of magnetic reconnection, inflows and outflows of fibrils are clearly detected \citep{2015ApJ...798L..11Y}. In another reconnection event, oscillation of newly reconnected fibrils after reconnection in the chromosphere is reported \citep{2016ApJ...819L..24Y}. Magnetic reconnection between the chromospheric fibrils and filament threads is studied, and suggested to play an important role in solar eruptions by releasing the magnetic twist \citep{2016NatCo...711837X, 2018ApJ...853L..26H}. Between two neighboring filaments, magnetic reconnection occurs, and forms two sets of new filaments \citep{2017ApJ...838..131Y} and two-sided-loop jets \citep{2019ApJ...883..104S, 2019ApJ...887..220Y}. Magnetic reconnection between the emerging and pre-existing fibrils is investigated \citep{2018ApJ...861..108Z, 2019ApJ...876...51Z}. An example of two-sided-loop jets simultaneously observed in the chromosphere, transition region, and corona is then described \citep{2018ApJ...861..108Z}. Recently, a small-scale oscillatory reconnection event is presented, that leads to the formation and disappearance of a flux rope \citep{2019ApJ...874L..27X}. Moreover, the disconnection of a filament caused by reconnection is revealed \citep{2020A&A...633A.121X}.

Magnetic reconnection modulated by non-local solar activities has been studied theoretically \citep[see a review by][]{2018SSRv..214...45M}. \citet{2006A&A...452..343N} numerically simulated the interaction of fast magnetoacoustic oscillations of a non-flaring loop with a nearby magnetic null point. They found that the fast magnetoacoustic wave coming into the null point from the outside oscillating loop can trigger the magnetic reconnection. \citet{2006SoPh..238..313C} performed magnetohydrodynamic (MHD) simulations of magnetic reconnection driven by five-minute solar p-mode oscillations. They pointed out that several typical and puzzling features of the transition-region explosive events can only be explained if there exist p-mode oscillations and the reconnection site is located in the upper chromosphere. \citet{2009A&A...493..227M} investigated the nature of nonlinear fast magnetoacoustic waves propagating in the neighborhood of a 2D magnetic X-point. They demonstrated that magnetic reconnection is naturally driven by the MHD wave propagation. However, magnetic reconnection affected by the external solar activities is rarely observed directly. In the wake of an erupting flux rope, oscillation of the current sheets caused by the neighboring filament eruption has been reported \citep{2016ApJ...829L..33L}. But the evolution of magnetic reconnection is not investigated. Recently, \citet{2017ApJ...851L...1Z} presented that the rising flux rope pushes the overlying loops, and forms an external current sheet where magnetic reconnection takes place. In this paper, we report a reconnection event accelerated by an adjacent filament eruption. The observations and results are shown separately in Sections\,\ref{sec:obs} and \ref{sec:res}. A summary and discussion is presented in Section\,\ref{sec:sum}.

\section{Observations}\label{sec:obs}

The NVST is a 1-meter ground-based solar telescope, located in the Fuxian Solar Observatory of the Yunnan Observatories, Chinese Academy of Sciences. It provides observations of the solar fine structures and their evolution in the solar lower atmosphere. On 2013 March 15, the NVST observed the active region (AR) 11696 with a field of view (FOV) of 200\arcsec$\times$186\arcsec~in the H$\alpha$ channel, centered at 6562.8\,\AA~with a bandwidth of 0.25 \AA, from 01:20\,UT to 06:40\,UT. The H$\alpha$ images have a time cadence of 12\,s and spatial sampling of 0.164\arcsec\,pixel$^{-1}$. They are processed first by flat field correction and dark current subtraction, and then reconstructed by speckle masking \citep[][and references therein]{2016NewA...49....8X}. The co-alignment of H$\alpha$ images is carried out by a fast sub-pixel image registration algorithm \citep{2012JKAS...45..167F, 2015RAA....15..569Y}.

The Atmospheric Imaging Assembly \citep[AIA;][]{2012SoPh..275...17L} onboard the Solar Dynamic Observatory \citep[SDO;][]{2012SoPh..275....3P} is a set of normal-incidence imaging telescopes, acquiring solar atmospheric images in ten wavelength bands. Different AIA channels show plasma at different temperatures, e.g., 131\,\AA~peaks at $\sim$10\,MK (Fe\,XXI) and  $\sim$0.6\,MK (Fe\,VIII), 94\,\AA~peaks at  $\sim$7.2\,MK (Fe\,XVIII), 335\,\AA~peaks at  $\sim$2.5\,MK (Fe\,XVI), 211\,\AA~peaks at  $\sim$1.9\,MK (Fe\,XIV), 193\,\AA~peaks at  $\sim$1.5\,MK (Fe\,XII), 171\,\AA~peaks at  $\sim$0.9\,MK (Fe\,IX), and 304\,\AA~peaks at  $\sim$0.05\,MK (He\,II). In this study, we employ the AIA images in one ultraviolet (UV) channel (1600\,\AA) and seven extreme UV (EUV) channels (131, 94, 335, 211, 193, 171, and 304\,\AA) to investigate the evolution of magnetic reconnection and filament eruption. Here, the AIA images are processed to 1.5-level using ``aia\_prep.pro". Then the spatial sampling of AIA images is 0.6\arcsec\,pixel$^{-1}$, and the time cadences of AIA EUV and UV images are 12\,s and 24\,s, respectively. The Helioseismic and Magnetic Imager \citep[HMI;][]{2012SoPh..275..229S} onboard the SDO provides line of sight (LOS) magnetograms, with a time cadence of 45\,s and spatial sampling of 0.5\arcsec\,pixel$^{-1}$. We use the HMI LOS magnetograms to study the evolution of surface magnetic fields underlying the reconnection region and the erupting filament. 

The NVST H$\alpha$ images have been rotated to match the orientation of SDO observations. All the data from different instruments, i.e., the SDO and NVST, and passbands have been aligned with a principle of best cross-correlation between images of two passbands with the closest characteristic temperatures. 

\section{Results}\label{sec:res}

\begin{figure}[ht!]
\centering
\includegraphics[width=0.7\textwidth]{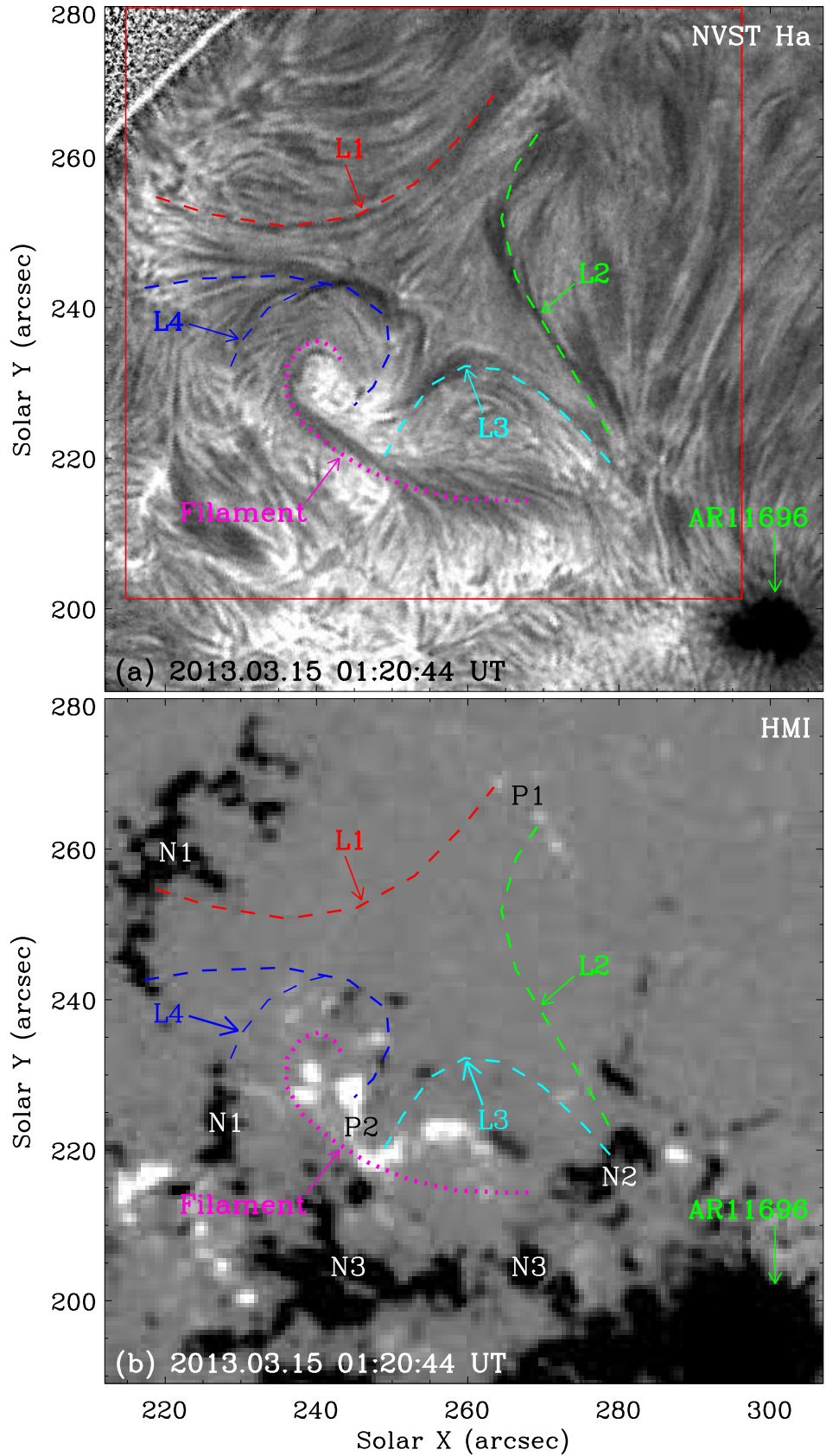}
\caption{General information of H$\alpha$ fibrils and their nearby filament. (a) NVST H$\alpha$ image and (b) SDO/HMI LOS  magnetogram. The red, green, cyan, and blue dashed lines outline the fibrils L1, L2, L3, and L4, respectively. The pink dotted lines represent the filament. The red rectangle in (a) shows the FOVs of Figures\,\ref{f:mr_ha}, \ref{f:mr_aia1}, \ref{f:filament_eruption}(a)-(f), and \ref{f:mr_aia2}. The N1, N2, and N3, and P1 and P2 in (b) separately denote the negative and positive magnetic fields. See Section\,\ref{sec:res} for details.
\label{f:general_information}}
\end{figure}

We analyze the observations that provide signatures of how a filament eruption causes magnetic reconnection to speed-up at a different location. On 2013 March 15, a saddle-like structure, located to the northeast of AR 11696, was observed by the NVST, see Figure\,\ref{f:general_information}(a). It is created by a set of four fibrils seen in the H$\alpha$ images, marked separately by L1, L2, L3, and L4, see the red, green, cyan, and blue dashed lines in Figure\,\ref{f:general_information}(a). A curved filament is located to the southeast of the saddle-like structure, outlined by the pink dotted line in Figure\,\ref{f:general_information}(a). We overlay the H$\alpha$ fibrils L1-L4 and their nearby filament on an HMI LOS magnetogram in Figure\,\ref{f:general_information}(b). For better descriptions, we label the positive and negative magnetic fields as P1 and P2, and N1, N2, and N3, based on the fibril connectivity, see Figure\,\ref{f:general_information}(b). The fibrils L1, L2, L3, and L4 connect the positive and negative magnetic fields P1 and N1, P1 and N2, P2 and N2, and P2 and N1, respectively, see the red, green, cyan, and blue dashed lines in Figure\,\ref{f:general_information}(b). The central positive magnetic fields P2 and their surrounding negative magnetic fields N1, N2, and N3 constitute a fan-spine magnetic field configuration, see Figures\,\ref{f:general_information}(b) and \ref{f:cartoon}. The filament is located upon the polarity inversion line between the central positive magnetic fields P2 and the surrounding negative magnetic fields N1 and N3, see the pink dotted line in Figure\,\ref{f:general_information}(b).

\subsection{Magnetic reconnection before the filament eruption}\label{sec:mr1}

\begin{figure}[ht!]
\centering
\includegraphics[width=0.8\textwidth]{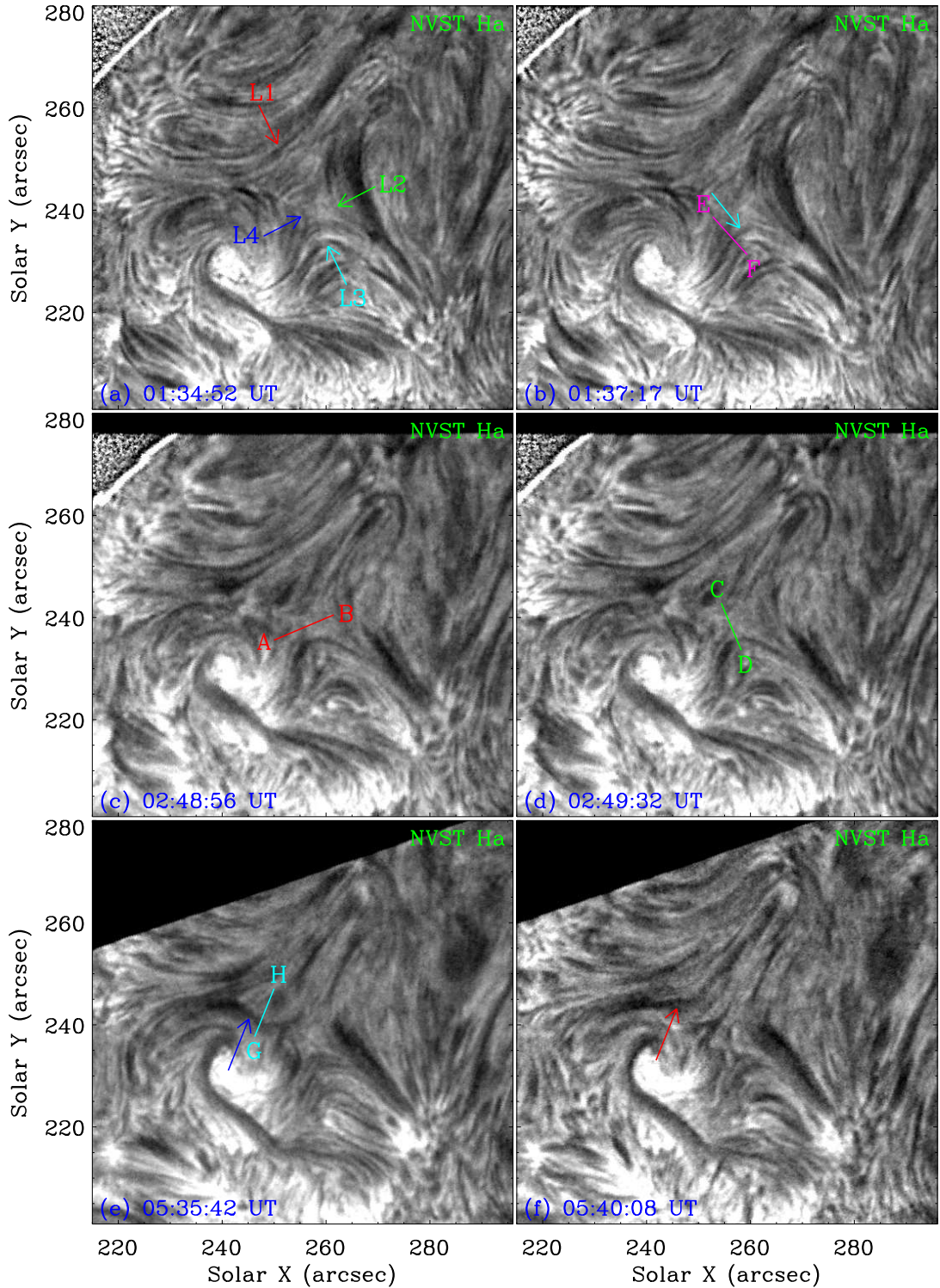}
\caption{Magnetic reconnection between H$\alpha$ fibrils observed by NVST. (a)-(f) NVST H$\alpha$ images. The red, green, cyan, and blue arrows separately denote the fibrils L1, L2, L3, and L4. The red, green, pink, and cyan lines AB, CD, EF, and GH in (c), (d), (b), and (e) show the positions for time slices of NVST H$\alpha$ images displayed in Figures\,\ref{f:measurements_ha}(a), (b), (c), and (d), respectively. The FOV is denoted by the red rectangle in Figure\,\ref{f:general_information}(a). An animation of the unannotated NVST H$\alpha$ images is available. It covers $\sim$5.3\,hr starting at 01:20:44 UT, and the video cadence is 12\,s. See Section\,\ref{sec:mr1} for details. (An animation of this figure is available.)
\label{f:mr_ha}}
\end{figure}

\begin{figure}[ht!]
\includegraphics[width=\textwidth]{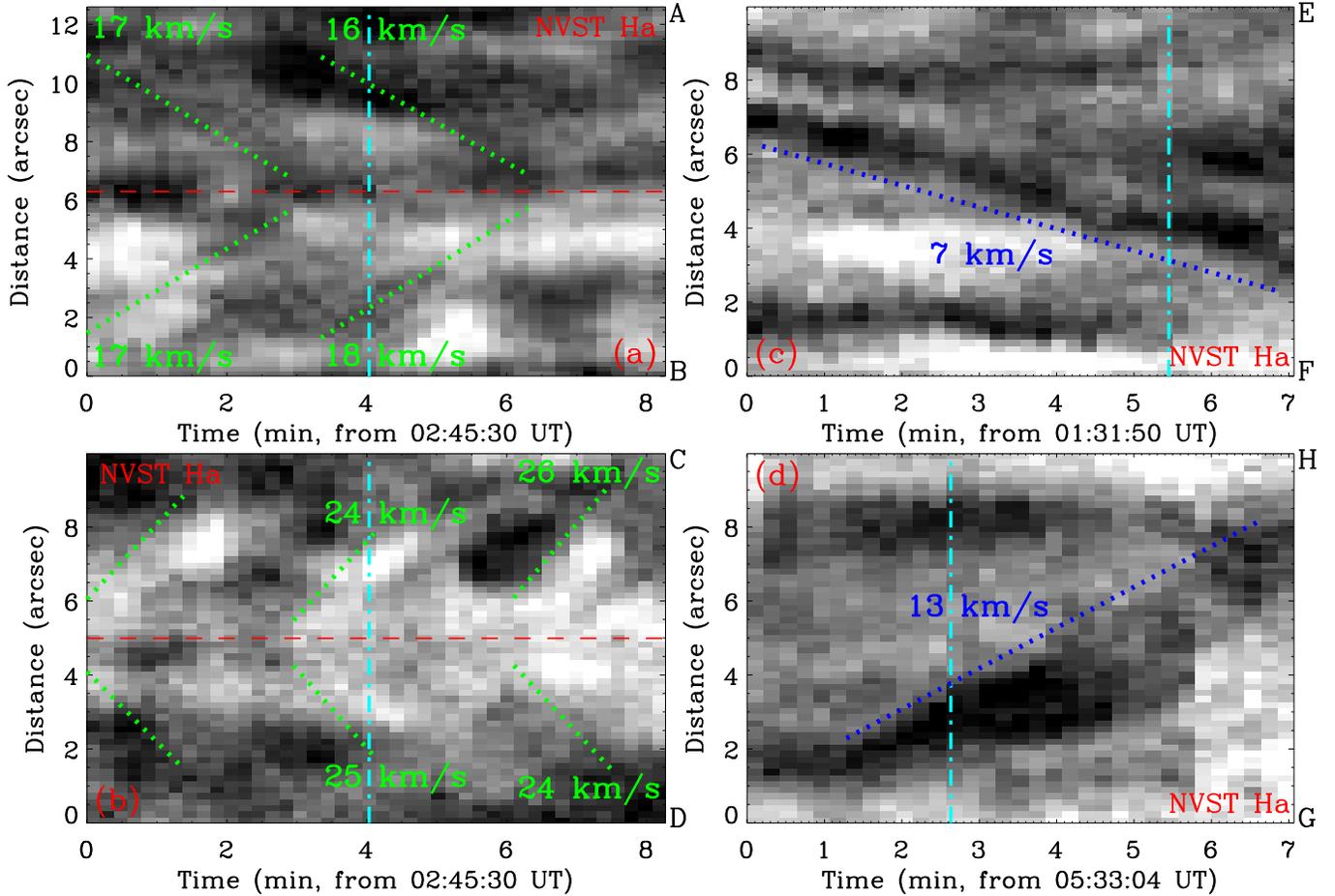}
\centering
\caption{Temporal evolution of the magnetic reconnection between H$\alpha$ fibrils observed by NVST. (a)-(d) Time slices of NVST H$\alpha$ images along the red, green, pink, and cyan lines AB, CD, EF, and GH in Figures\,\ref{f:mr_ha}(c), (d), (b), and (e), respectively. The green and blue dotted lines outline the motions of fibrils, with the moving speeds denoted by the numbers. The red dashed lines in (a)-(b) mark the reconnection region. The cyan dash-dotted lines in (a)-(d) separately show the times of H$\alpha$ images shown in Figures\,\ref{f:mr_ha}(c), (d), (b), and (e). See Section \ref{sec:mr1} for details.
\label{f:measurements_ha}}
\end{figure}

The H$\alpha$ fibrils L2 and L4 constitute a saddle-type structure, see Figure\,\ref{f:mr_ha}(a). They move toward each other, and reconnect. Two sets of newly reconnected fibrils L1 and L3 then form, and retract away from the reconnection region, see the online animated version of Figure\,\ref{f:mr_ha}. No current sheet is observed in the H$\alpha$ diagnostics in the reconnection region between fibrils L2 and L4. Along the red line AB in Figure\,\ref{f:mr_ha}(c), a time slice of H$\alpha$ images is made, and displayed in Figure\,\ref{f:measurements_ha}(a). Inward motions of fibrils L2 and L4, with mean speeds of $\sim$16-18 km\,s$^{-1}$, see the green dotted lines in Figure\,\ref{f:measurements_ha}(a), toward the reconnection region, denoted by the red dashed line in Figure\,\ref{f:measurements_ha}(a), are clearly identified. Along the green line CD in Figure\,\ref{f:mr_ha}(d), another time slice of H$\alpha$ images is obtained, and shown in Figure\,\ref{f:measurements_ha}(b). At the same time, outward motions of the newly reconnected fibrils L1 and L3, with mean speeds of $\sim$24-26 km\,s$^{-1}$, see the green dotted lines in Figure\,\ref{f:measurements_ha}(b), away from the reconnection region, denoted by the red dashed line in Figure\,\ref{f:measurements_ha}(b), are evidently detected. Moreover, topological reconfiguration of the fibrils by reconnection, e.g., from fibrils L4 to L3 and L1, is observed, see Figures\,\ref{f:mr_ha}(a)-(b) and (e)-(f) and the online animated version of Figure\,\ref{f:mr_ha}. Along the pink and cyan lines EF and GH in Figures\,\ref{f:mr_ha}(b) and (e), time slices of H$\alpha$ images are made, and illustrated in Figures\,\ref{f:measurements_ha}(c) and (d), respectively. The fibrils L4 reconnect with L2, and then separately turn into L3 and L1, with mean moving speeds of $\sim$7 km\,s$^{-1}$ and $\sim$13 km\,s$^{-1}$, see the blue dotted lines in Figures\,\ref{f:measurements_ha}(c)-(d).

The process of magnetic reconnection is also observed by AIA, see Figure\,\ref{f:mr_aia1}. Four sets of coronal loops L1-L4 are recorded in AIA EUV images, forming a saddle-like structure, consistent with the H$\alpha$ fibrils L1-L4, see Figure\,\ref{f:mr_aia1}(a). In the same manner, the loops L2 and L4 constitute an X-type structure, see the red dotted lines in Figure\,\ref{f:mr_aia1}(c). At the interface of these two loops, magnetic reconnection takes place, see the online animated version of Figure\,\ref{f:mr_aia2}. Along the cyan line IJ in Figure\,\ref{f:mr_aia1}(a), a time slice of AIA 304\,\AA~images is made, and displayed in Figure\,\ref{f:measurements_aia1}(a). Different from the H$\alpha$ observations, motions of the EUV loops toward the reconnection region are hard to observe, see the online animated version of Figure\,\ref{f:mr_aia2}. Nevertheless, the inward motion of loops L2, similar to the H$\alpha$ fibrils L2, with a mean speed of $\sim$33 km\,s$^{-1}$, see the blue dotted line in Figure\,\ref{f:measurements_aia1}(a), toward the reconnection region, marked by the purple dashed line in Figure\,\ref{f:measurements_aia1}(a), is still identified. 

In the reconnection region, a current sheet forms, denoted by green solid arrows in Figure\,\ref{f:mr_aia1}. In the particular snapshot, i.e., at 01:33:11 UT, it has a width of $\sim$2\,Mm, and a length of $\sim$7.2\,Mm in AIA 171\,\AA~images, see Figure\,\ref{f:mr_aia1}(c). Here, the length of current sheet is measured between two cusp-shaped structures at the ends of current sheet, marked by red pluses in Figure\,\ref{f:mr_aia1}(c). For the width of current sheet, first we get the intensity profile in the AIA 171 \AA~channel perpendicular to the current sheet, e.g., along the purple line in Figure\,\ref{f:mr_aia1}(c). Employing the intensity surrounding the current sheet, we calculate the background emission, and subtract it from the intensity profile. We fit the residual intensity profile using a single Gaussian, and obtain the full width at half maximum (FWHM) of the single Gaussian fit as the current sheet width. The magnetic reconnection rate, the ratio of the width and the length of the current sheet \citep{1957JGR....62..509P}, is thus $\sim$0.28. The current sheet appears in most AIA EUV channels, except the higher-temperature channels, e.g., 335 and 94\,\AA, see Figure\,\ref{f:mr_aia1} and the online animated version of Figure\,\ref{f:mr_aia2}. In the blue rectangle enclosing the current sheet in Figure\,\ref{f:mr_aia1}(e), the light curves of the AIA 304, 131, 171, 193, and 211\,\AA~channels are calculated, and shown in Figure\,\ref{f:measurements_aia1}(c). Considering the influence of background emission, similar temporal evolution of the AIA EUV light curves are identified. They reach the peaks at almost the same time, see the blue vertical dashed line in Figure\,\ref{f:measurements_aia1}(c). The current sheet repeatedly appears and disappears, see the online animated version of Figure\,\ref{f:mr_aia2}, with a mean period of $\sim$8\,minutes, obtained from the wavelet analysis of the AIA 304\,\AA~light curve in Figure\,\ref{f:measurements_aia1}(c). Same as before, the well-developed current sheets are measured using AIA 171\,\AA~images. They have the width of 1-2\,Mm with a mean value of 1.5\,Mm, the length of 5.3-7.2\,Mm with a mean value of 6.4\,Mm, and the reconnection rate of 0.18-0.3 with a mean value of 0.24. 

As the current sheet is not observed in the AIA higher-temperature channels, e.g., 94 and 335\,\AA, plasma at that location could be cooler than $\sim$2.5\,MK, the characteristic temperature of AIA 335\,\AA~channel. The current sheet in AIA 131\,\AA~images, see Figure \ref{f:mr_aia1}(b), hence shows plasma with the lower characteristic temperature ($\sim$0.6\,MK) of AIA 131\,\AA~channel. Using six AIA EUV channels, including 94, 335, 211, 193, 171, and 131\,\AA, we analyze the temperature and emission measure (EM) of the current sheet. Here, we employ the differential EM (DEM) analysis using ``xrt\_dem\_iterative2.pro" \citep{2012ApJ...761...62C}. The current sheet region, enclosed by the red rectangle in Figure\,\ref{f:mr_aia1}(b), is chosen to compute the DEM. The region out of the current sheet, enclosed by the purple rectangle in Figure\,\ref{f:mr_aia1}(b), is chosen for the background emission that is subtracted from the current sheet region. In each region, the DN counts in each of the six AIA channels are temporally normalized by the exposure time and spatially averaged over all pixels. The DEM curve of the current sheet region is displayed in Figure\,\ref{f:measurements_aia1}(e). Consistent with the AIA imaging observations, the DEM shows a lack of hot plasma component in the current sheet, see the black curve in Figure\,\ref{f:measurements_aia1}(e). The average DEM-weighed temperature and EM are 2\,MK and 2.5$\times$10$^{27}$\,cm$^{-5}$, respectively. 

Using the EM, the number density ($n_{p}$) of current sheet is estimated using $n_{p}=\sqrt{\frac{{\textrm{EM}}}{D}}$, where $D$ is the LOS depth of current sheet. Assuming that the depth $D$ equals the width ($W$) of current sheet, then the density is $n_{p}=\sqrt{\frac{{\textrm{EM}}}{W}}$. Employing EM=2.5$\times$10$^{27}$\,cm$^{-5}$ and W=2\,Mm, we obtain the density to be 3.5$\times$10$^{9}$\,cm$^{-3}$. The thermal energy (TE) of current sheet is also calculated using TE=$\frac{3}{2}$n$_{p}$$\cdot$k$_{B}$$\cdot$V$\cdot$$\delta$T. Here k$_{B}$ is Boltzmann's constant, V volume, and $\delta$T temperature increase from the temperature (T$_{1}$) of reconnection inflowing structure to that (T$_{2}$) of current sheet. Assuming that the current sheet is a cylinder, its volume is then V=$\pi$($\frac{W}{2}$)$^{2}$$\cdot$L, where L is the length of current sheet. Therefore the thermal energy is TE=$\frac{3}{2}$n$_{p}$$\cdot$k$_{B}$$\cdot$$\pi$($\frac{W}{2}$)$^{2}$$\cdot$L$\cdot$(T$_{2}$-T$_{1}$). As the reconnection inflow is observed mainly in H$\alpha$ images and the current sheet is identified in EUV images, we obtain T$_{1}$=10$^{4}$ K, the temperature of H$\alpha$ fibrils \citep{2012ApJ...749..136L}, and T$_{2}$ to be the average DEM-weighed temperature of current sheet. Employing n$_{p}$=3.5$\times$10$^{9}$\,cm$^{-3}$, W=1-2\,Mm, L=5.3-7.2\,Mm, T$_{2}$=2\,MK, and T$_{1}$=10$^{4}$ K, we calculate the thermal energy of current sheet, and get TE=(1.9$\pm$1.3)$\times$10$^{25}$\,erg.

Plasmoids appear in the current sheet, and propagate along it bi-directionally, and then further along the reconnection loops L2 and L4, see Figure \ref{f:mr_aia1}(d) and the online animated version of Figure\,\ref{f:mr_aia2}. Same as the current sheet, they are not observed in the AIA higher-temperature channels, e.g., 335 and 94\,\AA. Along the current sheet, see the green line KL in Figure\,\ref{f:mr_aia1}(c), a time slice of AIA 171\,\AA~images is made, and shown in Figure\,\ref{f:measurements_aia1}(b). It indicates that the plasmoids always form in the middle of the current sheet, marked by the red dashed line in Figure\,\ref{f:measurements_aia1}(b), and then move bi-directionally with mean speeds of $\sim$36-48\,km\,s$^{-1}$, see the green dotted lines in Figure\,\ref{f:measurements_aia1}(b). Along the blue line MN in Figure\,\ref{f:mr_aia1}(c), another time slice of AIA 171\,\AA~images is made, and illustrated in Figure\,\ref{f:measurements_aia1}(d). Several motions of plasmoids along the north leg of loops L2 are evidently identified, with a mean speed of $\sim$71\,km\,s$^{-1}$, see the green dotted line in Figure\,\ref{f:measurements_aia1}(d). 

From $\sim$02:30 UT, a set of higher-lying loops L6, connecting the surrounding negative magnetic fields N1-N3 and the remote positive magnetic fields, appears in most AIA EUV channels, except 304\,\AA, see Figure\,\ref{f:filament_eruption}(a). They are likely to be heated by nanoflares \citep{2015A&A...583A.109L}, as no significant activity is detected associated with the brightening of loops. The current sheet, i.e., the reconnection region, is then covered by the loops L6 in these EUV channels. It, however, still appears in AIA 304\,\AA~images, see the online animated version of Figure\,\ref{f:mr_aia2}. This indicates that the reconnection between loops L2 and L4 continues to take place as before, consistent with the H$\alpha$ observations. A set of lower-lying loops L5 overlying the filament is also detected. It connects the positive and negative magnetic fields P2 and N3,  see Figures\,\ref{f:filament_eruption}(a) and \ref{f:cartoon}. In the AIA higher-temperature channels, e.g., 94 and 335\,\AA, the loops L5 and L6 gradually disappear, indicating the cooling process of heated loops \citep{2015A&A...583A.109L}. Consistent with the previous observations, the current sheet forms in the lower-temperature, e.g., 304\,\AA, rather than the higher-temperature channels, e.g., 94 and 335\,\AA, see the online animated version of Figure\,\ref{f:mr_aia2}.

\subsection{Filament eruption}\label{sec:filament_eruption}

\begin{figure}[ht!]
\includegraphics[width=0.66\textwidth]{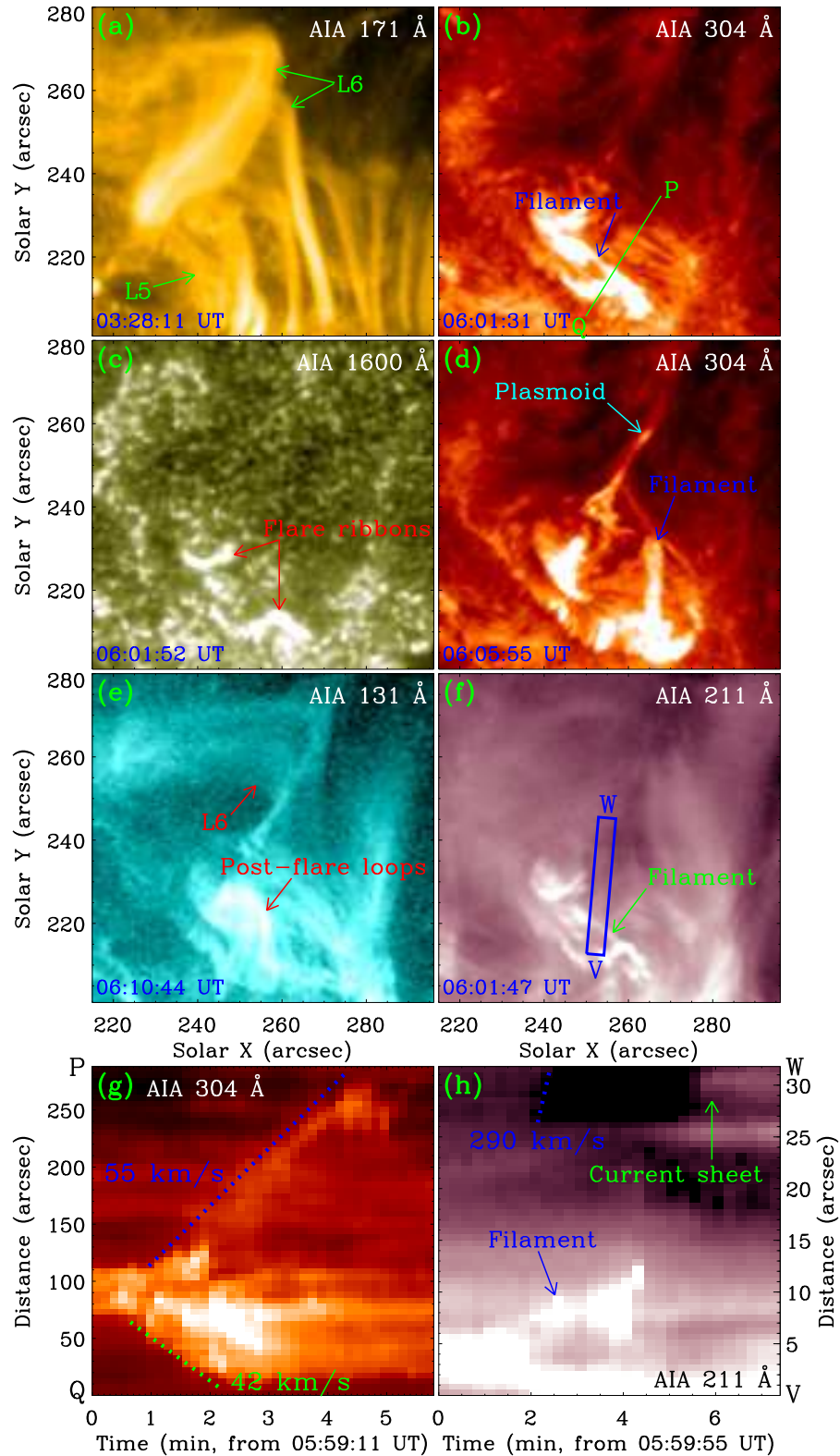}
\centering
\caption{Filament eruption observed by AIA. (a) AIA 171 \AA, (b) and (d) 304 \AA, (c) 1600 \AA, (e) 131 \AA, and (f) 211 \AA~images. (g)-(h) Time slices of AIA 304\,\AA~and 211\,\AA~images along the green line PQ in (b) and along the VW direction in the blue rectangle in (f), respectively. The blue and green dotted lines in (g) separately outline the filament eruption and flare ribbon separation. The blue dotted line in (h) outlines the motion of loops. The moving speeds are denoted by the numbers in (g)-(h). The FOVs of (a)-(f) are denoted by the red rectangle in Figure\,\ref{f:general_information}(a). See Sections\,\ref{sec:mr1} and \ref{sec:filament_eruption} for details. 
\label{f:filament_eruption}}
\end{figure}

From $\sim$05:52 UT, the north, rather than the south, part of the filament, located to the southeast of the reconnection region, brightens, see Figure\,\ref{f:filament_eruption}(b), and then erupts, see Figure\,\ref{f:filament_eruption}(d). A partial eruption of the filament is thus observed \citep{2016NatPh..12..847L}. Along the erupting direction, see the green line PQ in Figure\,\ref{f:filament_eruption}(b), a time slice of AIA 304\,\AA~images is made, and displayed in Figure\,\ref{f:filament_eruption}(g). The filament erupts with a mean projection speed of $\sim$55 km\,s$^{-1}$, see the blue dotted line in Figure\,\ref{f:filament_eruption}(g). Assuming that the filament erupts outward along the radial direction, we obtain the corrected erupting speed to be 163 km\,s$^{-1}$, using the heliographic position N13\,W15 of the erupting filament. The erupting filament is prevented eventually by the higher-lying loops L5 and L6, showing a failed filament eruption, and makes the higher-lying loops bright, see Figure\,\ref{f:filament_eruption}(e) and the online animated version of Figure\,\ref{f:mr_aia2}. Moreover, the failed filament eruption may also be caused by the reconnection between the erupting filament and its overlying loops \citep{2016NatPh..12..847L, 2020ApJ...889..106Y}. Two flare ribbons and post-flare loops, associated with the filament eruption, appear, see Figures\,\ref{f:filament_eruption}(c) and (e). The south flare ribbon moves away from the polarity inversion line of the positive and negative magnetic fields P2 and N3, with a mean speed of $\sim$42 km\,s$^{-1}$, see the green dotted line in Figure\,\ref{f:filament_eruption}(g). 

The filament eruption pushes away the higher-lying loops L6 covering the reconnection region, see Figure\,\ref{f:filament_eruption}(f) and the online animated version of Figure\,\ref{f:mr_aia2}. It thus leads to a disturbance propagating outward across the reconnection region. Along the propagating direction VW in the blue rectangle in Figure\,\ref{f:filament_eruption}(f), a time slice of AIA 211\,\AA~images is measured, and shown in Figure\,\ref{f:filament_eruption}(h). It indicates that the loops L6 is pushed away with a mean moving speed of $\sim$290\,km\,s$^{-1}$ by the propagating disturbance caused by the filament eruption, see the blue dotted line in Figure\,\ref{f:filament_eruption}(h). A dimming region then forms, and the reconnection region, e.g., the current sheet, reappears in these AIA EUV channels, e.g., 171, 193, and 211\,\AA, see the online animated version of Figure\,\ref{f:mr_aia2}. 

\subsection{Magnetic reconnection after the filament eruption}\label{sec:mr2}

After the filament eruption, the current sheet between loops L2 and L4 has the width and length of $\sim$2\,Mm and $\sim$3.5\,Mm at 06:18:35 UT, and thus a reconnection rate of $\sim$0.57, see Figure\,\ref{f:mr_aia2}(c). Here, the width is similar to the current sheet widths (1-2\,Mm), the length is, however, smaller than the current sheet lengths (5.3-7.2\,Mm), and the reconnection rate is much larger than the reconnection rates (0.18-0.3), before the filament eruption. The current sheet, different from those occurring in the AIA low-temperature EUV channels before the filament eruption, appears in all AIA EUV channels, marked by the green solid arrows in Figure\,\ref{f:mr_aia2}. Same as in Section\,\ref{sec:mr1}, in the blue rectangle enclosing the reconnection region in Figure\,\ref{f:mr_aia2}(e), the light curves of the AIA 94, 335, 211, 193, 171, 131, and 304\,\AA~channels are calculated, and displayed in Figure\,\ref{f:measurements_aia2}(a). Here, the AIA 211, 193, and 171\,\AA~light curves before the filament eruption, denoted by the red vertical dotted line in Figure\,\ref{f:measurements_aia2}(a), are not measured, because before the filament eruption the reconnection region is covered by the higher-lying loops L6 in these channels, see Section\,\ref{sec:mr1}. Before the filament eruption, the AIA 94\,\AA~and 335\,\AA~light curves keep constant, indicating that no current sheet appears in these AIA higher-temperature channels. The AIA 304\,\AA~and 131\,\AA~light curves, however, evolve due to the appearance of current sheet in these lower-temperature channels. Right after the filament eruption, the AIA 335, 211, 193, and 171\,\AA~light curves decrease evidently, showing a dimming, as the higher-lying loops L6, covering the reconnection region, are pushed away by the propagating disturbance caused by the filament eruption, see also Figures\,\ref{f:filament_eruption}(f) and (h). All the light curves then exhibit rapid rise and reach the peaks. Among them, the AIA 94\,\AA~light curve reaches the peak at 06:09:37 UT, see the green vertical dashed line in Figure\,\ref{f:measurements_aia2}(a), $\sim$2.3 minutes later than the AIA 304\,\AA~light curve that peaks at 06:07:19 UT, see the blue vertical dashed line in Figure\,\ref{f:measurements_aia2}(a). Affected by the brightening of higher-lying loops caused by the filament eruption, see Section \ref{sec:filament_eruption}, the AIA 335 \AA~light curve reaches the peak several minutes later than the other light curves. It, however, has a small peak at 06:07:26 UT, identical to the peak of the AIA 304 \AA~light curve, see Figure\,\ref{f:measurements_aia2}(a). In addition, a much bright reconnection region is detected simultaneously in H$\alpha$ observations, see the online animated version of Figure\,\ref{f:mr_ha}.

As the current sheet appears in all AIA EUV channels, it, therefore, may contain plasma with both of the high and low temperature. The current sheet region, enclosed by the red rectangle in Figure\,\ref{f:mr_aia2}(d), is chosen to compute the DEM. The region out of the current sheet, enclosed by the purple rectangle in Figure\,\ref{f:mr_aia2}(d), is selected for the background emission that is subtracted from the current sheet region. The DEM curve of the current sheet region is shown in Figure\,\ref{f:measurements_aia2}(c). It indicates that more plasma with higher temperature and less plasma with lower temperature is detected in the current sheet,  comparing to that in Figure\,\ref{f:measurements_aia1}(e). The average DEM-weighed temperature and EM are 5.5\,MK and 1.7$\times$10$^{28}$\,cm$^{-5}$, respectively. Both these quantities are significantly higher than those of current sheet before the filament eruption (2\,MK and 2.5$\times$10$^{27}$\,cm$^{-5}$). Similar to Section\,\ref{sec:mr1}, the density of current sheet is also estimated to be 9.2$\times$10$^{9}$\,cm$^{-3}$ by using the EM, under the assumption that the LOS depth (D) of the current sheet equals its width (W=2\,Mm). It is larger than that of current sheet before the filament eruption (3.5$\times$10$^{9}$\,cm$^{-3}$). Using n$_{p}$=9.2$\times$10$^{9}$\,cm$^{-3}$, W=2\,Mm, L=3.5\,Mm, T$_{2}$=5.5\,MK, and T$_{1}$=10$^{4}$ K, we also calculate the thermal energy of current sheet, and obtain the value to be TE=1.1$\times$10$^{26}$\,erg. It is also larger than that of current sheet before the filament eruption (1.9$\pm$1.3$\times$10$^{25}$\,erg).

In the current sheet, plasmoids appear and move along it and the reconnection loops, see Figure \ref{f:filament_eruption}(d) and the online animated version of Figure\,\ref{f:mr_aia2}. The north endpoint of loops L2, enclosed by the pink circles, then brightens, see Figure\,\ref{f:mr_aia2}(h). Along the green line RS in Figure\,\ref{f:mr_aia2}(c), a time slice of AIA 171\,\AA~images is made, and displayed in Figure\,\ref{f:measurements_aia2}(b). It shows that more plasmoids moving along the north leg of loops L2 are detected, comparing with that in Figure\,\ref{f:measurements_aia1}(d), with a similar mean speed of $\sim$70\,km\,s$^{-1}$, see the green dotted line in Figure\,\ref{f:measurements_aia2}(b). Moreover, the plasmoids, different from those appearing in the AIA lower-temperature EUV channels before the filament eruption, see Section\,\ref{sec:mr1}, appear in all AIA EUV channels. More hotter plasmoids are thus generated in the current sheet after the filament eruption. Comparing all the results before and after the filament eruption, see Sections\,\ref{sec:mr1} and \ref{sec:mr2}, we conclude that the reconnection between loops L2 and L4 is significantly accelerated by the filament eruption occurring to the southeast of the reconnection region.

\section{Summary and discussion}\label{sec:sum}

\begin{figure}[ht!]
\includegraphics[width=\textwidth]{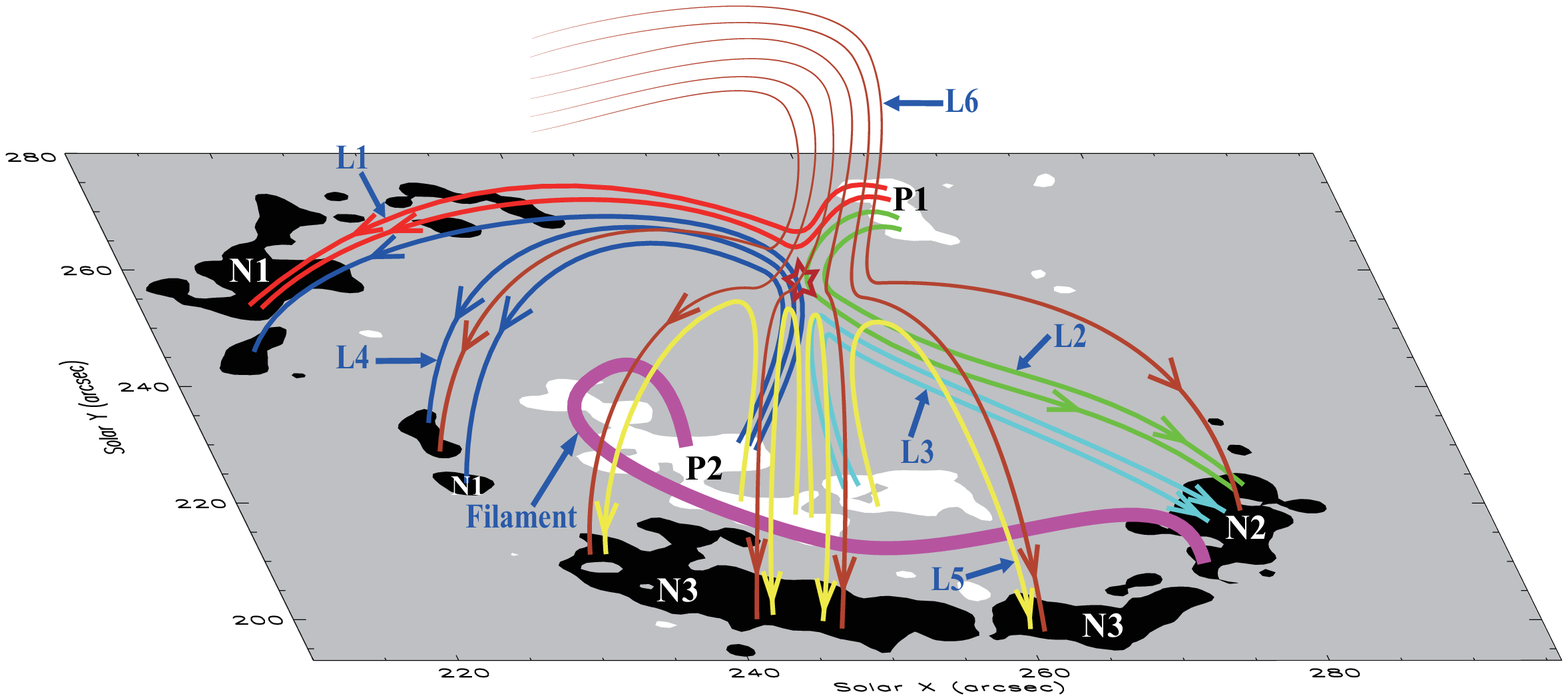}
\centering
\caption{Schematic diagram of the magnetic reconnection between loops (fibrils) and its nearby filament eruption. The red, green, cyan, blue, yellow, and brown lines with arrows represent the directional magnetic field lines of loops (fibrils) L1, L2, L3, L4, L5, and L6, respectively. The pink line shows the filament. The red star denotes the reconnection point between field lines L2 and L4. The gray parallelogram represents the photosphere, and the black and white patches separately indicate the negative and positive magnetic fields N1, N2, and N3, and P1 and P2. See Section \ref{sec:sum} for details.
\label{f:cartoon}}
\end{figure}

Employing the H$\alpha$ images from NVST, and the AIA images and HMI LOS magnetograms from SDO, we study the reconnection between fibrils (loops) L2 and L4, and its nearby filament eruption. The reconnection accelerated by the filament eruption is then reported. In H$\alpha$ images, a saddle-like structure, consisting of four sets of fibrils L1-L4, is observed. The fibrils L2 and L4 from opposite sides of the saddle region move together, and reconnect. The newly reconnected fibrils L1 and L3 then form, and retract away from the reconnection region. In AIA EUV images, similar loops L1-L4 and their evolution are identified. At the interface of loops L2 and L4, the current sheet repeatedly forms and disappears. Magnetic reconnection takes place in the current sheet. Plasmoids appear in the current sheet, and propagate along it, and then further along the reconnection loops. A filament, located to the southeast of the reconnection region, partially erupts, and leads to a flare. It is then prevented by the overlying loops as a failed filament eruption. After the filament eruption, a hotter, shorter current sheet forms with a much larger reconnection rate, where more hotter plasmoids appear. Based on the NVST H$\alpha$ images, and the SDO AIA EUV images and HMI LOS magnetograms, a schematic diagram of the magnetic reconnection between fibrils (loops) and its nearby filament eruption is demonstrated in Figure\,\ref{f:cartoon}. Here, the red star represents the reconnection point between magnetic field lines of loops L2 and L4.

A small-scale reconnection event among a saddle-like structure is observed by NVST. Similar to the small-scale reconnection events previously reported \citep{2015ApJ...798L..11Y, 2016ApJ...819L..24Y}, inward and outward motions of H$\alpha$ fibrils toward and away from the reconnection region are evidently detected, see Section\,\ref{sec:mr1}. The reconnection inflowing and outflowing speeds of $\sim$17 km\,s$^{-1}$ and $\sim$25 km\,s$^{-1}$ are consistent with those of the fast reconnection event in \citet{2015ApJ...798L..11Y}. Different from \citet{2015ApJ...798L..11Y}, in this study the current sheet appears only in the AIA EUV channels, rather than the H$\alpha$ channel, see Section\,\ref{sec:mr1}. This indicates that the current sheet is significantly heated during the reconnection process \citep{2016NatPh..12..847L, 2016ApJ...829L..33L, 2020A&A...633A.121X}. The width (1-2\,Mm) and length (3.5-7.2\,Mm) of current sheets are identical to those in \citet{2016NatCo...711837X, 2020A&A...633A.121X}, but larger than those in \citet{2015ApJ...798L..11Y} and \citet{2018ApJ...858L...4X}. Moreover, the reconnection rate (0.18-0.57) is similar to those in \citet{2016NatCo...711837X, 2018ApJ...858L...4X}, but larger than those in \citet{2020A&A...633A.121X}. Many plasmoids form in the current sheet, suggesting the presence of plasmoid instabilities during the process of magnetic reconnection \citep{2013A&A...557A.115K, 2016NatPh..12..847L, 2019A&A...628A...8P}. They propagate along the current sheet bi-directionally, and then further along the reconnection loops with a mean speed of $\sim$70 km\,s$^{-1}$. The moving speed here is consistent with those in \citet{2015ApJ...798L..11Y} and \citet{2020A&A...633A.121X}, but smaller than those in \citet{2016NatPh..12..847L}.

\begin{figure}[ht!]
\includegraphics[width=0.8\textwidth]{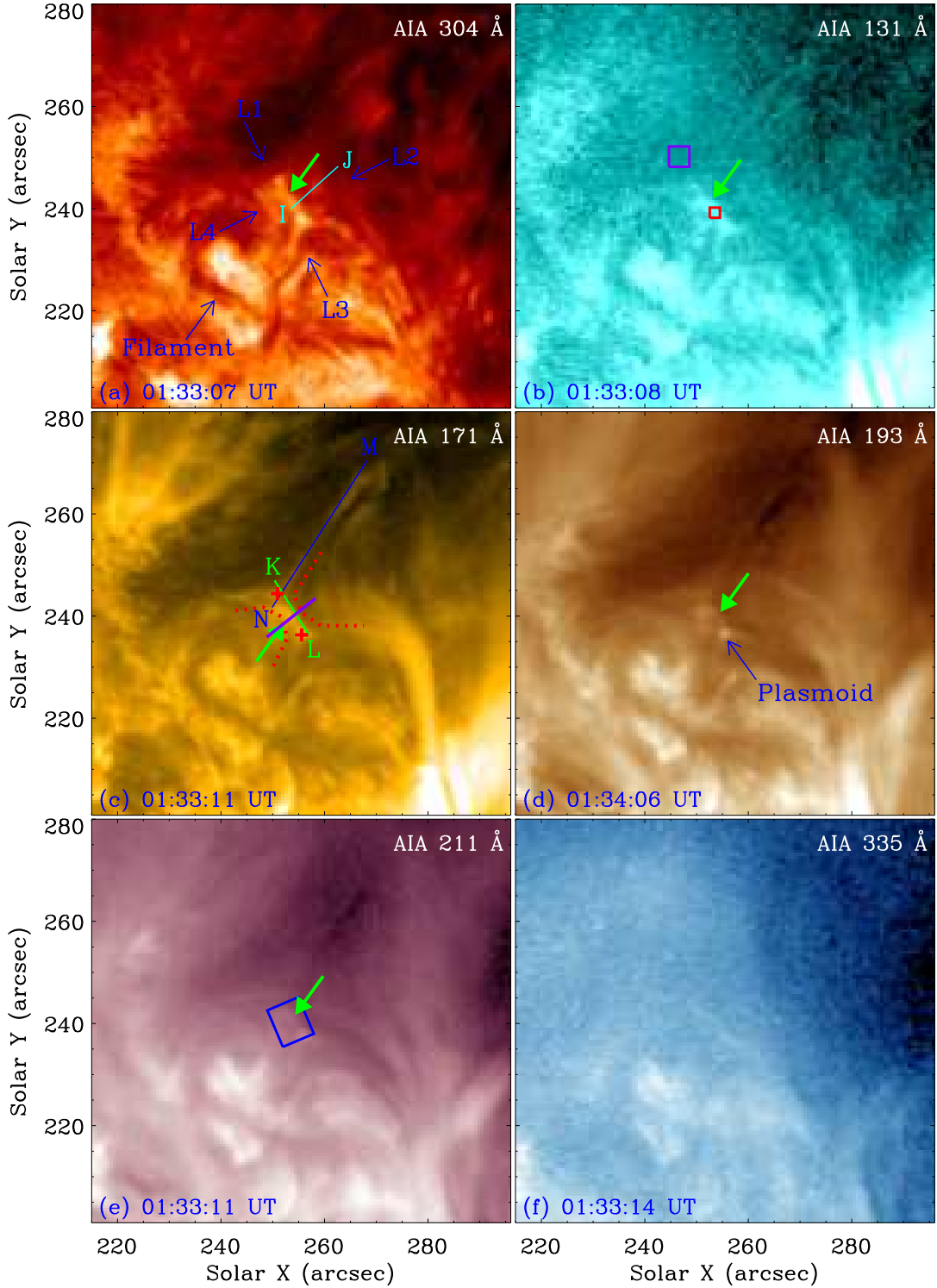}
\centering
\caption{Magnetic reconnection between loops before the filament eruption observed by SDO/AIA. (a) AIA 304 \AA, (b) 131 \AA, (c) 171 \AA, (d) 193 \AA, (e) 211 \AA, and (f) 335 \AA~images. The green solid arrows in (a)-(e) denote the current sheet. The cyan, green, and blue lines IJ, KL, and MN in (a) and (c) show the positions for time slices of AIA 304 \AA~and 171 \AA~images displayed in Figures\,\ref{f:measurements_aia1}(a), (b), and (d), respectively. The red and purple rectangles in (b) separately enclose the region for the DEM curve in Figure\,\ref{f:measurements_aia1}(e), and the location where the background emission is measured. In (c), the red pluses mark the positions between which the length of current sheet is measured, the purple line denotes the position along which the width of current sheet is calculated, and the red dotted lines outline the loops L2 and L4. The blue rectangle in (e) marks the region for the light curves of the AIA EUV channels as shown in Figure\,\ref{f:measurements_aia1}(c). The FOV is denoted by the red rectangle in Figure\,\ref{f:general_information}(a). See Section\,\ref{sec:mr1} for details.
\label{f:mr_aia1}}
\end{figure}

\begin{figure}[ht!]
\includegraphics[width=0.9\textwidth]{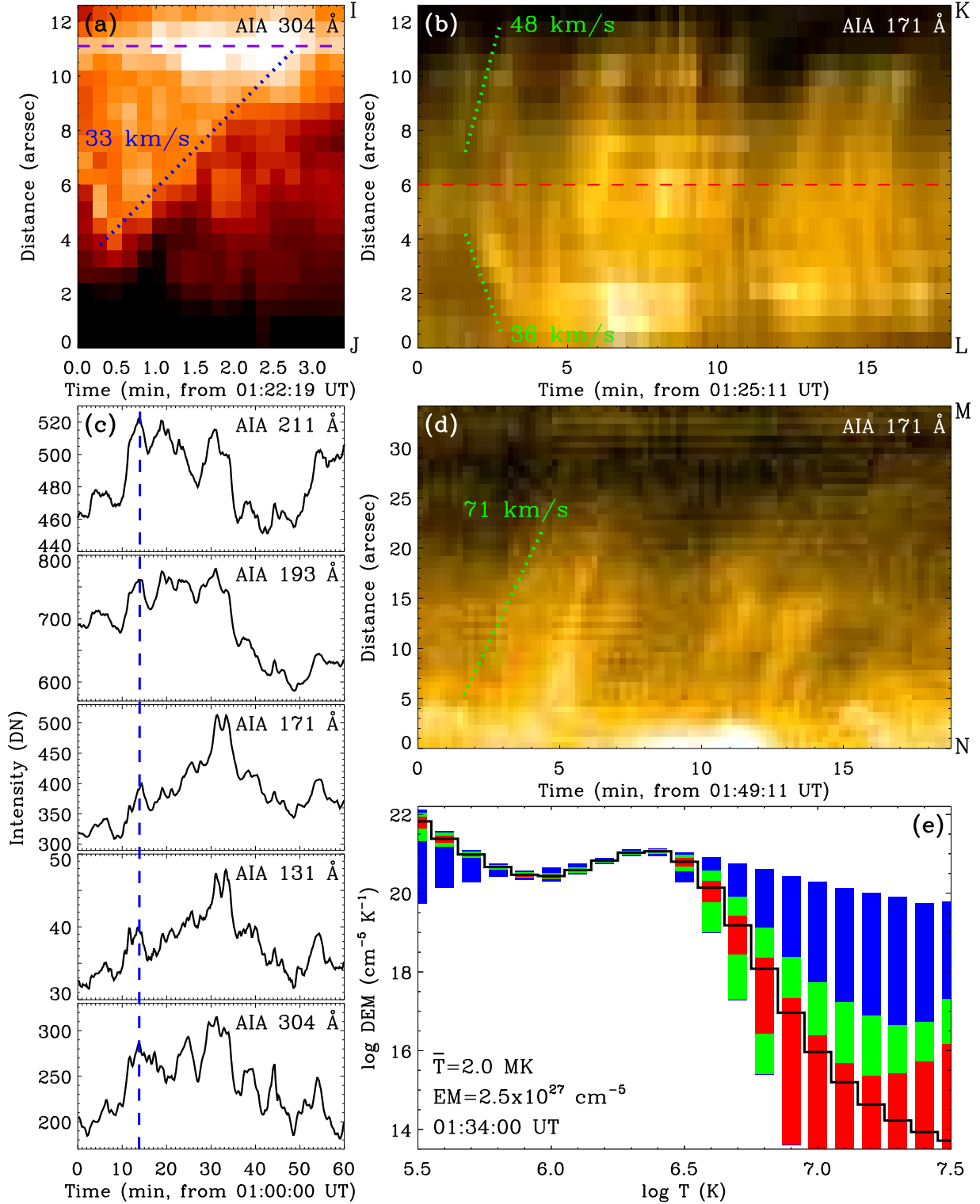}
\centering
\caption{Temporal evolution of the magnetic reconnection between loops before the filament eruption observed by AIA. Time slices of (a) AIA 304\,\AA, and (b) and (d) 171\,\AA~images along the cyan, green, and blue lines IJ, KL, and MN in Figures\,\ref{f:mr_aia1}(a) and (c), respectively. (c) Light curves of the AIA EUV channels in the blue rectangle in Figure\,\ref{f:mr_aia1}(e). (e) DEM curve for a current sheet region enclosed by the red rectangle in Figure\,\ref{f:mr_aia1}(b). The blue and green dotted lines in (a), (b), and (d) separately outline the motions of loops and plasmoids, with moving speeds denoted by the numbers. The purple dashed line in (a) denotes the reconnection region. The red dashed line in (b) marks the middle of the current sheet. The blue vertical dashed line in (c) marks a peak of the EUV light curves. In (e), the black curve is the best-fit DEM distribution, and the red, green, and blue rectangles separately represent the regions containing 50\%, 51-80\%, and 81-95\% of the Monte Carlo solutions. See Section \ref{sec:mr1} for details.
\label{f:measurements_aia1}}
\end{figure}

\begin{figure}[ht!]
\includegraphics[width=0.66\textwidth]{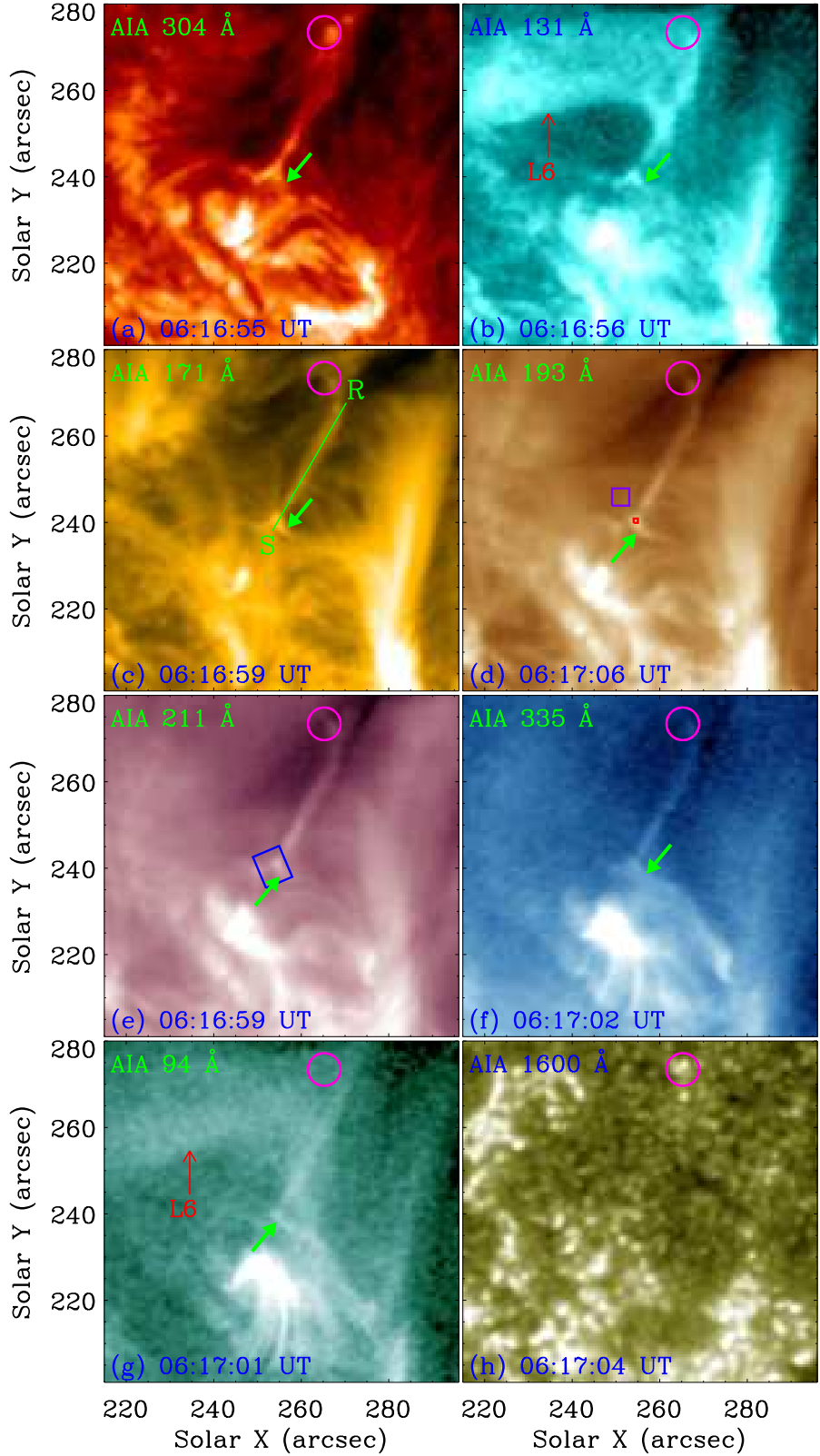}
\centering
\caption{Magnetic reconnection between loops after the filament eruption observed by AIA. (a) AIA 304 \AA, (b) 131 \AA, (c) 171 \AA, (d) 193 \AA, (e) 211 \AA, (f) 335 \AA, (g) 94 \AA, and (h) 1600 \AA~images. The pink circles enclose the north endpoint of loops L2. The green solid arrows in (a)-(g) denote the current sheet. The green line RS in (c) shows the position for time slice of AIA 171 \AA~images as displayed in Figure\,\ref{f:measurements_aia2}(b). The red and purple rectangles in (d) separately enclose the region for the DEM curve in Figure\,\ref{f:measurements_aia2}(c), and the location where the background emission is measured. Same as in Figure\,\ref{f:mr_aia1}(e), the blue rectangle in (e) marks the region for the light curves of the AIA EUV channels as shown in Figure\,\ref{f:measurements_aia2}(a). The FOV is denoted by the red rectangle in Figure\,\ref{f:general_information}(a). An animation of the unannotated AIA images is available. It covers $\sim$5.4 hr starting at 01:00 UT, and the video duration is 12\,s. See Section \ref{sec:mr2} for details. (An animation of this figure is available.)  
\label{f:mr_aia2}}
\end{figure}

\begin{figure}[ht!]
\includegraphics[width=0.8\textwidth]{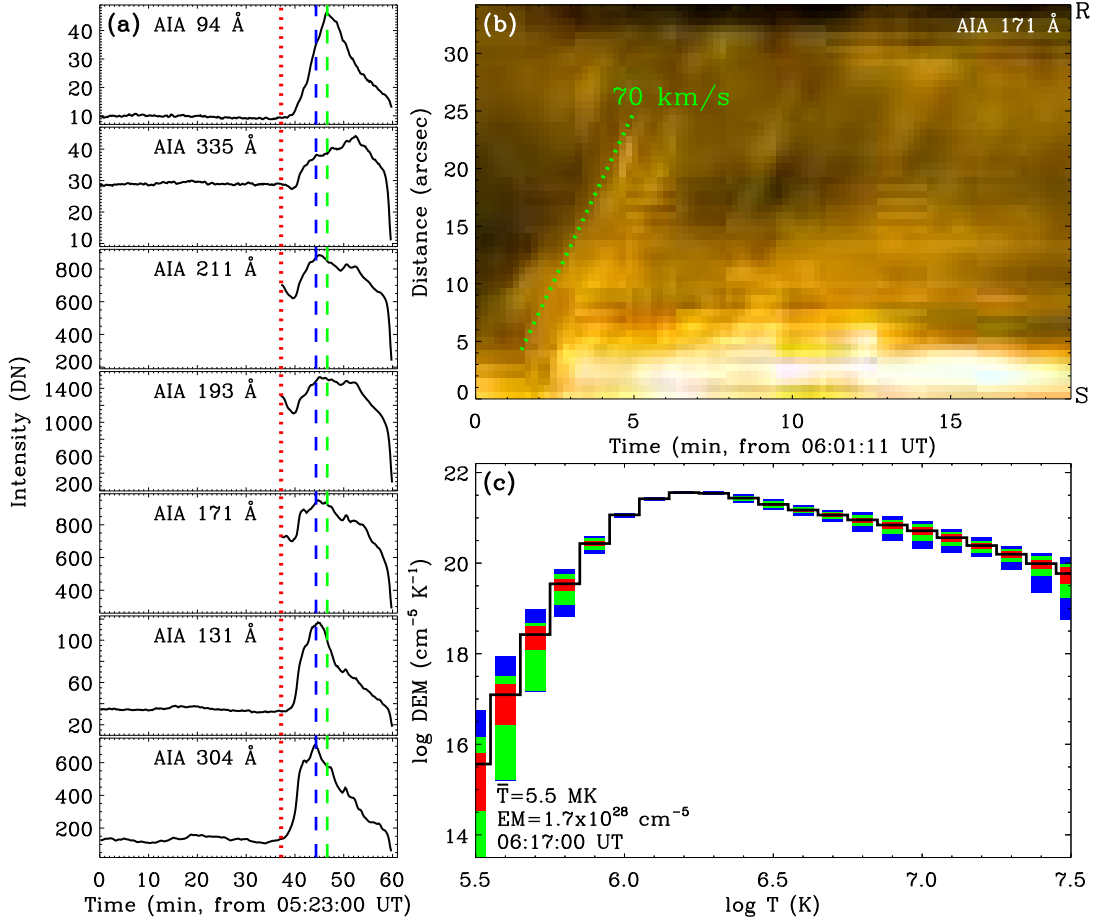}
\centering
\caption{Temporal evolution of the magnetic reconnection between loops after the filament eruption observed by AIA. (a) Light curves of the AIA EUV channels in the blue rectangle in Figure\,\ref{f:mr_aia2}(e). (b) Time slice of AIA 171 \AA~images along the green line RS in Figure\,\ref{f:mr_aia2}(c). (c) DEM curve for a current sheet region enclosed by the red rectangle in Figure\,\ref{f:mr_aia2}(d). In (a), the red vertical dotted line denotes the filament eruption, and the blue and green vertical dashed lines separately mark the peaks of the AIA 304 \AA~and 94 \AA~light curves. The green dotted line in (b) outlines the motion of plasmoids, with moving speed denoted by the number. In (c), the black curve is the best-fit DEM distribution, and the red, green, and blue rectangles represent the regions containing 50\%, 51-80\%, and 81-95\% of the Monte Carlo solutions, respectively. See Section \ref{sec:mr2} for details. 
\label{f:measurements_aia2}}
\end{figure}

Magnetic reconnection accelerated by nearby filament eruption is observed. After the filament eruption, the length of current sheet decreases significantly from 5.3-7.2\,Mm to 3.5\,Mm. The reconnection rate, however, increases largely from 0.18-0.3 to 0.57, see Sections\,\ref{sec:mr1} and \ref{sec:mr2}. The enhancements of the AIA EUV light curves in the reconnection region after the filament eruption, see Figures\,\ref{f:mr_aia1}(c) and \ref{f:mr_aia2}(a), suggest the increase of temperature and/or density of plasma in the current sheet. The current sheet appears in the AIA higher-temperature channels, e.g., 335 and 94\,\AA, after rather than before the filament eruption. It is thus heated to much higher temperature after the filament eruption. This is also supported by the DEM curves of current sheet before and after the filament eruption, e.g., the average DEM-weighed temperature of current sheet increases from 2\,MK to 5.5\,MK. The increase of current sheet density from 3.5$\times$10$^{9}$ cm$^{-3}$ to 9.2$\times$10$^{9}$ cm$^{-3}$ after the filament eruption shows that more plasma (n$_{p}$$\cdot$$\pi$$\cdot$($\frac{W}{2}$)$^{2}$$\cdot$L) from (4.7$\pm$3.2)$\times$10$^{34}$ to 1$\times$10$^{35}$ is heated to higher temperature. More thermal energy of current sheet converted by reconnection after (1.1$\times$10$^{26}$\,erg) than before ((1.9$\pm$1.3)$\times$10$^{25}$\,erg) the filament eruption is then achieved. In addition, more hotter plasmoids form in the current sheet after the filament eruption during the same time intervals, see Figures\,\ref{f:measurements_aia1}(d) and \ref{f:measurements_aia2}(b). This indicates that more plasma is accelerated during the reconnection process. More kinetic energy is hence converted by reconnection after the filament eruption.

Magnetic reconnection may be accelerated by the fast MHD wave caused by filament eruption. \citet{2006A&A...452..343N} suggested that the fast wave coming into the magnetic null point from the outside leads to the increase of electric current density. The increasing electric current then efficiently induces plasma micro-instabilities of various kinds, and hence produces anomalous resistivity, which efficiently triggers the reconnection. Fast mode MHD waves generated by filament eruptions are indeed widely reported \citep[e.g.,][]{2010ApJ...723L..53L, 2012A&A...539A...7L, 2018ApJ...853....1S}. In this study, the higher-lying loops covering the reconnection region in the AIA EUV channels are pushed away right after the filament eruption, see Section\,\ref{sec:filament_eruption}. This suggests that the filament eruption leads to a disturbance propagating outward across the reconnection region, that could be related to fast mode MHD wave. Using v$_{A}$=$\frac{B}{\sqrt{4 \pi n_{p} \cdot m_{p}}}$, we calculate the Alfv$\acute{e}$n speed near the reconnection region, where B is the magnetic field strength, and m$_{p}$ is the proton mass. Employing the magnetic field strength B=(3$\pm$2) G in the corona \citep{2020Sci...369..694Y}, and the coronal density n$_{p}$=(1$\pm$0.5)$\times$10$^{9}$ cm$^{-3}$, less than the current sheet density (3.5$\times$10$^{9}$ cm$^{-3}$) before the filament eruption, the Alfv$\acute{e}$n speed is obtained to be v$_{A}$=(272$\pm$216) km\,s$^{-1}$. It is consistent with the propagating speed ($\sim$290 km\,s$^{-1}$) of the disturbance. This propagating disturbance is hence likely to represent the fast-mode MHD wave driven by the filament eruption. The fast wave comes into the current sheet, increases its electric current density \citep{2006A&A...452..343N},  and accelerates the magnetic reconnection. It could also enhance turbulent plasma motions at the current sheet, leading to a turbulent reconnection \citep[][]{2020ApJ...890L...2C}. Additionally, the propagating disturbance may also push more magnetic flux of loops (fibrils) L4, and thus more magnetic energy, into the reconnection region \citep{2017ApJ...851L...1Z}. All these effects will play a role in liberating magnetic energy with a larger reconnection rate. The released magnetic energy will then be converted to other forms of energy, e.g., thermal and kinetic.

\acknowledgments
The authors thank the referee for helpful comments. We are indebted to the NVST and SDO teams for providing the data. This work is supported by the Strategic Priority Research Program of Chinese Academy of Sciences, Grant No. XDB 41000000, the National Natural Science Foundations of China (12073042, 11673034, 11533008, 11790304, 11873059, 1111903050, and 11773039), and the Key Research Program of Frontier Sciences (ZDBS-LY-SLH013) and the Key Programs (QYZDJ-SSW-SLH050) of Chinese Academy of Sciences. We acknowledge the usage of JHelioviewer software \cite[][]{2017A&A...606A..10M} and NASA's Astrophysics Data System.  


\end{document}